\documentclass[useAMS,usenatbib]{mn2e}
\usepackage{times}
\usepackage{epsfig}

\newcommand{\ebv}{E(B$-$V)}

\title[An Asssessment of Broadband Optical Colours as Age Indicators for Star Clusters]{An Asssessment of Broadband Optical Colours as Age Indicators for Star Clusters}

\author[M. Hancock, B. J. Smith, M. L. Giroux, and C. Struck]{M. Hancock$^{1}$\thanks{E-mail:
hancockm@etu.edu (MH); smithbj@etsu.edu (BJS); girouxm@etsu.edu (MLG); curt@iastate.edu (CS)},  B. J. Smith$^{1}$, M. L. Giroux$^{1}$, and C. Struck$^{2}$\\
$^{1}$Department of Physics, Astronomy, and Geology, East Tennessee State University, Box 70652, Johnson City, TN 37614\\
$^{2}$Department of Physics and Astronomy, Iowa State University, Ames IA 50011}

\begin{document}

%\date{Accepted 1988 December 15. Received 1988 December 14; in original form 1988 October 11}

\pagerange{\pageref{firstpage}--\pageref{lastpage}} \pubyear{2008}

\maketitle

\label{firstpage}

\begin{abstract}
We present an empirical assessment of the use of broadband optical
colours as age indicators for unresolved extragalactic clusters and
investigate stochastic sampling effects on integrated colours.   We
use the integrated properties of Galactic open clusters as models for
unresolved extragalactic clusters.  The population synthesis code {\it
Starburst99} \citep{lei99} and four optical colours were used to
estimate how well we can recover the ages of 62 well-studied Galactic
open clusters with published ages.   We provide a method for
estimating the ages of unresolved clusters and for reliably
determining the uncertainties in the age estimates.  Our results
support earlier conclusions based on comparisons to synthetic
clusters, namely the (U$-$B) colour is critical to the estimation of
the ages of star forming regions.    We compare the observed optical
colours with those obtained from {\it Starburst99} using the published
ages and get good agreement.  The scatter in the
(B$-$V)$_{observed}-$(B$-$V)$_{model}$ is larger for lower luminosity
clusters, perhaps due to stochastic effects.
\end{abstract}


\section{INTRODUCTION}

Thousands of luminous young star clusters have been discovered in
external galaxies by the Hubble Space Telescope (HST) (e.g.,
\citealp{hol92,hol96,whi93,whi95,meu95,joh99,ben02,kee03,han03,whi03,wei04},
and references therein).  Population synthesis analyses show that
these clusters are sometimes very young, with ages of a few to a few
tens of Myr.  Such age information can be extremely valuable, because
it provides clues to cluster formation processes as well as cluster
destruction mechanisms. A radial gradient in the ages of clusters
within a galaxy can tell us about gas inflow driven by a bar or an
interaction.  For interacting galaxies, comparison of cluster ages
with dynamical models can provide information about star forming
mechanisms. For example, gas compression along the Arp 107 tidal arm
may have caused a gradient of ages along the arm (e.g.,
\citealp{smi05a}), while in M51, a burst of cluster formation occurred
at around the time of the latest passage of the companion
(\citealp{bas05,lee05}).  Cluster synthesis studies have led to the
suggestion that there is a high `infant mortality' in star clusters,
with many dissolving within 10 Myr (e.g., \citealp{bas05,fal05}).
Star clusters are themselves sometimes clustered into complexes with
characteristic radii of $\sim$1 kpc (e.g., \citealp{zha01,lar04}), and
the more massive clusters tend to be located  near the centre of these
complexes, suggesting cluster merging.  Accurate age dating of young
clusters associated with `ultraluminous X-ray sources' (ULXs) can help
distinguish between stellar-mass and intermediate-mass (100$-$1000
M$_{\sun}$) black holes for the origin of the X-ray emission (e.g.,
\citealp{smi05b}).

Stellar population synthesis models have evolved dramatically in the
last several  decades, since the pioneering works where data on
globular clusters and individual giant stars were used to fill gaps in
the evolutionary tracks to make early colour models (e.g.\
\citealp{tin68,tin72,str78}).  Recently, several groups have
introduced evolutionary synthesis codes, e.g., \citet{bru93} (B\&C),
\citet{fri94} (GALEV), \citet{fio97} (PEGASE), and \citet{lei99}
(Starburst99).  These new models vary in terms of their input physics,
stellar spectral libraries and extinction laws.  In general, however,
there is good agreement among these models (e.g., \citealp{cha96,vaz05}).

Unfortunately, however, at the present time, it is not clear how
accurate age estimates based on these models are for unresolved
extragalactic clusters.  In most cases spectra of the targets are not
available, so the amount of extinction and metal abundance are
unknown.  In these cases, one typically compares some combination of
UV, optical, and/or near-IR broadband colours to model cluster spectral
energy distributions (SED) to simultaneously estimate age and
extinction assuming some metallicity (e.g. \citealp{pas03,han03,han07,smi08}).

There are several parameters that can affect the colour of a cluster
other than age.  Reddening plays a very important role, as does the
chemical composition.  However, in some age ranges and colours,
reddening and chemical composition are degenerate with age.
Furthermore, for low mass clusters ($\la10^{5}$ M$_{\odot}$), the
observed integrated colours  can be affected by stochastic sampling of
the initial mass function (IMF) (see for example,
\citealp{cer06,cer04,cer03}  and references therein).  For example,
the random addition of a small number of high mass stars will affect
the integrated colour of a cluster.  This suggests that the observed
integrated colours of low mass  clusters may not be good indicators of
age when compared to the integrated colours of a stellar  population
model with a fully sampled IMF.

It is not clear how well a particular set of colours
can predict the age of a cluster.  Are some colour sets better suited
than others?  What uncertainties can be expected because of the choice
of colours in the comparison?  What uncertainties can be expected
because of the assumptions made in generating the model SEDs?

Several authors have investigated the use of broadband colours as age
indicators by comparison of model colours to the colours of synthetic
clusters (e.g. \citealp{gil02,and04,deg05}).   We use an alternative
method by comparing model colours to the integrated colours of
resolved Galactic open clusters (OCs).  We compare the published
integrated colours of well-studied OCs to a set of population
synthesis models.  These clusters have published ages previously
determined by the turn-off of the zero-age main sequence on the H-R
diagram.  This work parallels that of \citet{pes08}, who do a similar
analysis of Magellanic Cloud clusters using both optical and near-IR
data.

The present paper is organized as follows.  In \S2 we describe our
sample of OCs.  We describe the population synthesis models used in
this study in \S3 and the data analysis in \S4.   Predicting the ages
of the OCs, predicting the amount of extinction and the effects of
metallicity, stochastic sampling effects and cluster dissolution
are discussed in \S5.  Finally, we summarize in \S6.

\section{THE DATA SAMPLE}

We started with the set of Galactic open clusters
from the WEBDA\footnote{http://www.univie.ac.at/webda/}(Web Base
Donn\'{e}es Amas) database operated at the Institute for Astronomy of
the University of Vienna \citep{mer95}.  The WEBDA database contains
379 OCs, with well-determined integrated (U$-$B), (B$-$V), (V$-$R),
and (V$-$I) colours, and colour excesses, as found by several authors
(\citealp{lat02, bat94, pan89, spa85, sag83, gra65}).  We then culled the
WEBDA sample to include only the sample of well-studied  OCs in
\citet{pau06}, who established a list of 72 open clusters  with
the most accurate known parameters to serve as a standard table for
testing isochrones and stellar models.    The age uncertainties in
\citet{pau06} were determined by measuring  the standard deviation of
all the published ages in the literature for each of the OCs.

Not all of the 4 integrated colours were determined for each of the
OCs in the \citet{pau06} standard set.  To model observations with
unknown dust extinction, we reversed the extinction corrections to the
published colours using the published values of \ebv\ and the
conversions from \ebv\ to the other colour excesses.  We used the
conversions in \citet{lat02}, namely E(U$-$B)=0.72 \ebv$+0.05$
\ebv$^2$, E(V$-$R)=0.6 \ebv, and E(V$-$I)=1.25 \ebv.

From the standard set we created sub-samples of OCs for each of the 4
optical colours and 10 different combinations of colours.   When
multiple integrated colour measurements were available,  we adopted
the most recently determined values.   Our final data sets include OCs
with mean age uncertainties of 19\% and ages ranging from 8 Myr to 8.8
Gyr.  The metallicities associated with this sample ranges from [Fe/H]
$\sim$1/7 solar to $\sim$2.3 solar.

Unfortunately, the WEBDA database does not give the uncertainties on
the total colours for individual OCs.  According to  \citet{sag83},
the maximum uncertainty in the published integrated colours for the
WEBDA data set is $\pm0.2$ mag.   One of the sources of uncertainty
listed is the error in the reddening.  Because we have used the
published colour excesses to reverse the extinction corrections, we
can neglect the extinction uncertainty in the colours.  Removing this
from the total uncertainty, assuming it was originally added in
quadrature, the maximum uncertainty in  colour is $\sim$0.14.  We
assume that all the measured colours have this maximum uncertainty.
This assumption further allows us to make fair comparisons of both 
the accuracy and precision afforded by each colour in age estimation.

\section{THE MODEL CLUSTERS}

We used a set of evolutionary synthesis models from the {\it
Starburst99} (SB99) code \citep{lei99}.  We used the new v5.1 code,
which includes the Padova asymptotic giant branch (AGB) stellar models
\citep{vaz05}.  The new version accounts for all stellar phases that
contribute to  the integrated light of a stellar population with
arbitrary age from extreme UV to NIR.  Strictly speaking, the Geneva
tracks are more appropriate for modeling  young clusters, less than 10
Myr, when O stars are present.  Most of our sample OCs have ages
greater than 100 Myr so for simplicity we only consider models with
the Padova tracks.

Our SB99 model spectral energy distributions (SEDs) were generated
assuming a Kroupa initial mass function (IMF) (favors high mass stars)
\citep{kro02} with exponents of 1.3 and 2.3 and mass ranges from 
$0.1-0.5$ M$_{\odot}$ and $0.5-100$ M$_{\odot}$ respectively.
We have also assumed instantaneous (single burst) star formation, and solar
abundances.  It has been demonstrated that adopting different forms of
the IMF has a minor impact on optical colours \citep{mac04}, so we
have not explored various IMFs.  We  also generated models  with
abundances less than (0.2$\times$) and greater than (2.5$\times$)
solar.   The model SEDs were reddened from 0.0 mag to 2.0 mag in 0.02
mag increments using the \citet{car89}  reddening law.  Finally, the
model SEDs were convolved with the Johnson and Kron-Cousins {\it
UBVRI} filter bandpasses and the broadband optical colours were
determined.  The models span a range of ages from 1 Myr to 20 Gyr.
From 1 Myr to 1 Gyr we used a step size of 1 Myr; from 1.1 to 20 Gyr,
the step size was 100 Myr.  Because the models will be compared to the
integrated colours of resolved stellar populations, only the stellar
contributions were included in the SEDs.  Nebular emission lines have
been shown to be important in the first $10^7$ yr \citep{and03} and
should be considered when studying unresolved stellar populations (see
\S5.3).

Figure 1 plots the (B$-$V) vs.\ (U$-$B) colour-colour diagram.
The black curves are our SB99 models with solar abundances and varying
E(B$-$V).  From the bottom to the top, E(B$-$V)=0.0, 0.5, 1.0, and 1.5
mags.  The filled squares are the observed colours of the standard set
of open clusters in the WEBDA sample.  For each curve, age increases
to the right from 1 Myr to 20 Gyr.  Note that we have assumed the
maximum uncertainty in colour for each OC.

\begin{figure}
\centering{\epsfig{figure=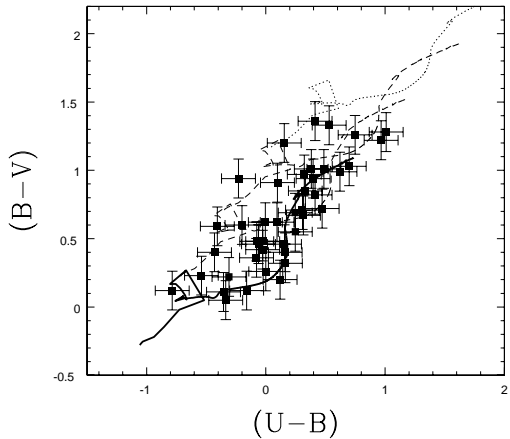, height=7.0cm}}
\caption[f1.eps]{(B$-$V) vs.\ (U$-$B) colour-colour diagram.  The black curves are our SB99 models with solar abundances and varying E(B$-$V).  From the bottom to the top, E(B$-$V)=0.0 (thick solid curve), 0.5 (dashed curve), 1.0 (dashed curve), and 1.5 (dotted curve) mags.  The filled squares are the observed colours of the standard set of open clusters in the WEBDA sample.  For each curve, age increases to the right from 1 Myr to 20 Gyr.  Note that we have assumed the maximum uncertainty in colour for each OC.}  \label{f1}
\end{figure}

\section{DATA ANALYSIS}

This study is not intended as a test of the SB99 model's ability to
reproduce the colours of young clusters.  This has already been
demonstrated (see for e.g., \citealp{vaz05}).  We intend to test  how
well various colours and initial model assumptions can recover the
ages of well studied Galactic open clusters.  In addition to using the
observed colours to estimate ages,  we compared the published colours
with the model colours calculated assuming the published ages to look
for evidence of stochastic sampling  effects.

To determine the model predicted ages of the OCs we compared the
observed integrated colours to each of the reddened (and unreddened)
SB99 model colours, for a single assumed metallicity.   We used a
$\chi^2$ minimization calculation (e.g., \citealp{pas03,deg05}) to determine
the best match of the observed colours to the models and hence the
ages of the open clusters:
\[
\chi^2 = \sum^{N}_{i=1}\left(\frac{obs_{i} - model_{i}}{\sigma_{i}}\right)^2
\]

\noindent where N is the number of colours (1-4) used in the analysis,
obs$_i$ is the observed colour,  model$_i$ is the corresponding model
colour, and $\sigma_i$ is the uncertainty in the obs$_i$ colour.  All
ages with a fit of $\chi^2\leq$ N  were considered good fits.  For
each colour set there was therefore a range of predicted ages for each
OC.   The age associated with the minimum $\chi^2$ is taken as  the
best-fit age.  To determine the uncertainties in the predicted age we
find the minimum and maximum ages within a $\Delta\chi^2$ defined to
give a 68\% confidence level (e.g., \citealp{pre92}).  Additionally,
we add the model step size to the age uncertainty.  We have also
predicted the amount of extinction in each case.  The amount of
extinction applied to the model associated with the best-fit age is
the best-fit \ebv.  The uncertainty in the predicted \ebv\ is
determined from the minimum and maximum \ebv\ within the same
$\Delta\chi^2$ mentioned above.  Additionally, we add the model step
size to the \ebv\ uncertainty.

\begin{table*}
\centering
\begin{minipage}{140mm}
\caption[ages colour differences]{Ages and Colour Differences} 
%\vspace{.2in} 
\begin{tabular}{lccrrrrrrr} 
\hline \hline 
Name & Age$^1$ &  Age$^2$  & M$_{V}$ & (U$-$B)  & (U$-$B) & diff. & (B$-$V)   & (B$-$V) & diff. \\ 
     & (pub)   &  (pred)    &         & measured & model   &       &  measured & model   &       \\
\hline 
Bochum 10& 8$\pm$2 & 1$_{-0}^{+31}$  & -6.52 & -1.05 & -0.67 & -0.38 & -0.24 & 0.27 & -0.51 \\ 
NGC 6871& 9$\pm$2 & 23$_{-22}^{+60}$  & -7.28 & -0.89 & -0.80 & -0.09 & -0.24 & 0.17 & -0.41 \\ 
Stock 14& 10$\pm$2 & 31$_{-30}^{+96}$  & -8.25 & -0.39 & -0.76 & 0.37 & 0.34 & 0.17 & 0.17 \\ 
King 12& 11$\pm$1 & 13$_{-12}^{+67}$  & -4.90 & -0.88 & -0.72 & -0.16 & -0.20 & 0.14 & -0.34 \\ 
IC 2581& 13$\pm$3 & 120$_{-114}^{+980}$  & -8.11 & -0.33 & -0.67 & 0.34 & 0.07 & 0.05 & 0.02 \\ 
NGC 3105& 21$\pm$3 & 4$_{-3}^{+49}$  & -6.95 & -0.69 & -0.67 & -0.02 & 0.11 & 0.06 & 0.05 \\ 
NGC 6250& 22$\pm$5 & 66$_{-61}^{+157}$  & -4.69 & -0.59 & -0.66 & 0.07 & -0.16 & 0.06 & -0.22 \\ 
NGC 6396& 30$\pm$8 & 38$_{-37}^{+2862}$  & -5.03 & -0.64 & -0.58 & -0.06 & -0.05 & 0.07 & -0.12 \\ 
Trumpler 1& 30$\pm$6 & 11$_{-10}^{+29}$  & -5.65 & -0.88 & -0.58 & -0.30 & -0.04 & 0.07 & -0.11 \\ 
Lynga 6& 35$\pm$9 & 30$_{-29}^{+103}$  & -4.38 & -0.65 & -0.54 & -0.11 & 0.01 & 0.07 & -0.06 \\ 
Melotte 20& 43$\pm$18 & 83$_{-77}^{+114}$  & -5.71 & -0.42 & -0.48 & 0.06 & 0.02 & 0.07 & -0.05 \\ 
King 10& 45$\pm$11 & 106$_{-100}^{+10594}$  & -5.34 & -0.58 & -0.47 & -0.11 & -0.19 & 0.07 & -0.26 \\ 
NGC 129& 62$\pm$15 & 9$_{-8}^{+20}$  & -6.03 & -0.64 & -0.40 & -0.24 & 0.39 & 0.09 & 0.30 \\ 
NGC 6834& 65$\pm$18 & 131$_{-125}^{+1269}$  & -6.20 & -0.45 & -0.39 & -0.06 & -0.10 & 0.10 & -0.20 \\ 
NGC 6405& 71$\pm$21 & 181$_{-134}^{+146}$  & -3.65 & -0.27 & -0.38 & 0.11 & -0.03 & 0.11 & -0.14 \\ 
Collinder 394& 74$\pm$6 & 97$_{-91}^{+178}$  & -4.74 & -0.26 & -0.38 & 0.12 & 0.23 & 0.11 & 0.12 \\ 
NGC 6025& 74$\pm$22 & 82$_{-76}^{+121}$  & -4.38 & -0.46 & -0.38 & -0.08 & -0.12 & 0.11 & -0.23 \\ 
NGC 5662& 77$\pm$20 & 78$_{-73}^{+1222}$  & -4.38 & -0.24 & -0.37 & 0.13 & 0.30 & 0.11 & 0.19 \\ 
NGC 6087& 78$\pm$19 & 143$_{-137}^{+309}$  & -5.09 & -0.21 & -0.37 & 0.16 & 0.18 & 0.12 & 0.06 \\ 
NGC 1778& 129$\pm$29 & 149$_{-143}^{+659}$  & -4.16 & -0.27 & -0.25 & -0.02 & 0.09 & 0.14 & -0.05 \\ 
NGC 1647& 130$\pm$25 & 111$_{-105}^{+197}$  & -2.35 & -0.35 & -0.25 & -0.10 & 0.07 & 0.14 & -0.07 \\ 
NGC 5316& 166$\pm$33 & 119$_{-113}^{+12681}$  & -4.91 & 0.13 & -0.19 & 0.32 & 0.67 & 0.15 & 0.52 \\ 
NGC 744& 184$\pm$49 & 256$_{-170}^{+1044}$  & -5.17 & -0.14 & -0.16 & 0.02 & 0.06 & 0.15 & -0.09 \\ 
Melotte 105& 224$\pm$53 & 281$_{-183}^{+1119}$  & -4.27 & -0.11 & -0.10 & -0.01 & 0.07 & 0.16 & -0.09 \\ 
NGC 3532& 262$\pm$46 & 243$_{-163}^{+363}$  & -5.50 & -0.03 & -0.04 & 0.01 & 0.22 & 0.18 & 0.04 \\ 
NGC 2548& 364$\pm$102 & 486$_{-341}^{+411}$  & -3.23 & 0.12 & 0.07 & 0.05 & 0.26 & 0.23 & 0.03 \\ 
Melotte 111& 522$\pm$82 & 384$_{-217}^{+273}$  & -1.95 & 0.12 & 0.16 & -0.04 & 0.20 & 0.33 & -0.13 \\ 
NGC 2527& 619$\pm$163 & 317$_{-204}^{+783}$  & -1.90 & 0.09 & 0.18 & -0.09 & 0.30 & 0.39 & -0.09 \\ 
NGC 2266& 736$\pm$77 & 498$_{-348}^{+1302}$  & -3.83 & 0.40 & 0.18 & 0.22 & 0.62 & 0.46 & 0.16 \\ 
Berkeley 2& 794$\pm$1 & 999$_{-953}^{+19101}$  & -4.18 & 0.14 & 0.17 & -0.03 & 0.46 & 0.49 & -0.03 \\ 
NGC 2355& 833$\pm$137 & 238$_{-187}^{+3562}$  & -3.67 & 0.22 & 0.17 & 0.05 & 0.59 & 0.51 & 0.08 \\ 
Berkeley 69& 867$\pm$48 & 506$_{-373}^{+19594}$  & -3.16 & 0.21 & 0.17 & 0.04 & 0.38 & 0.52 & -0.14 \\ 
NGC 2477& 875$\pm$238 & 250$_{-180}^{+10350}$  & -5.70 & 0.17 & 0.17 & 0.00 & 0.49 & 0.53 & -0.04 \\ 
NGC 2192& 1072$\pm$48 & 1000$_{-956}^{+2200}$  & -3.53 & 0.10 & 0.17 & -0.07 & 0.49 & 0.62 & -0.13 \\ 
NGC 2660& 1351$\pm$291 & 694$_{-585}^{+19406}$  & -4.08 & 0.34 & 0.22 & 0.12 & 0.61 & 0.72 & -0.11 \\ 
NGC 1798& 1421$\pm$16 & 47$_{-44}^{+9653}$  & -4.86 & 0.15 & 0.25 & -0.10 & 0.82 & 0.77 & 0.05 \\ 
NGC 2506& 1648$\pm$485 & 259$_{-175}^{+2941}$  & -4.31 & 0.22 & 0.28 & -0.06 & 0.54 & 0.82 & -0.28 \\ 
NGC 7044& 1824$\pm$361 & 19999$_{-19810}^{+102}$  & -3.93 & 0.48 & 0.26 & 0.22 & 0.58 & 0.81 & -0.23 \\ 
Berkeley 32& 3477$\pm$698 & 4100$_{-4094}^{+8600}$  & -2.97 & 0.28 & 0.36 & -0.08 & 0.78 & 0.91 & -0.13 \\ 
NGC 6253& 3949$\pm$1086 & 6800$_{-6794}^{+10399}$  & -2.68 & 0.34 & 0.38 & -0.04 & 0.81 & 0.92 & -0.11 \\ 
NGC 2682& 4093$\pm$958 & 1300$_{-1294}^{+8200}$  & -3.16 & 0.28 & 0.38 & -0.10 & 0.78 & 0.92 & -0.14 \\ 
NGC 6791& 7850$\pm$2026 & 19999$_{-19805}^{+102}$  & -4.14 & 0.82 & 0.46 & 0.36 & 1.02 & 0.97 & 0.05 \\ 
\hline \\ 
\multicolumn{10}{l}{\footnotesize{$^1$ Ages in Myr from \citet{pau06}}}\\
\multicolumn{10}{l}{\footnotesize{$^2$ Ages in Myr predicted from (U$-$B) and (B$-$V)}}\\
\end{tabular} 
\end{minipage}
\end{table*}

In Table 1, we list the subset of our sample that has the {\it U},
{\it B},  and {\it V} magnitudes available.  Column one is the name of
the OC, column two is the published age, column three is the predicted
age determined from the (U$-$B) and (B$-$V), column four is M$_{V}$,
columns five and six are the reddening-corrected measured and zero
reddened model (U$-$B) colours respectively, column seven is the
difference of the measured and model (U$-$B), and columns eight, nine
and ten are the same but for the (B$-$V) colour.  It can be seen from
Table 1 that for 74\% of the OCs, the (U$-$B) differences are
$\la0.14$, the  assumed maximum measurement uncertainty in measured
colour, while 60\% of the (B$-$V) differences are $\la0.14$.

\section{RESULTS}

\subsection{Predicting Ages}

In Figure 2, we plot the predicted ages against the published ages, if
only (U$-$B) and (B$-$V) are used to estimate the ages.  We also provide
a histogram showing the distribution of the differences between the
predicted and published ages, a plot of the predicted E(B$-$V)s against the
published E(B$-$V)s, and a histogram showing the distribution of the
differences between the predicted and published E(B$-$V)s.
In this Figure, we used the solar abundance models.  Plots of the
predictions determined with  the other colour combinations and at other
assumed metallicities are similar so are not shown.  
We see from this figure that with the combination of (U$-$B) and (B$-$V)
the ages of the OCs can be predicted with reasonable accuracy.

\begin{figure*}
\centering{\epsfig{figure=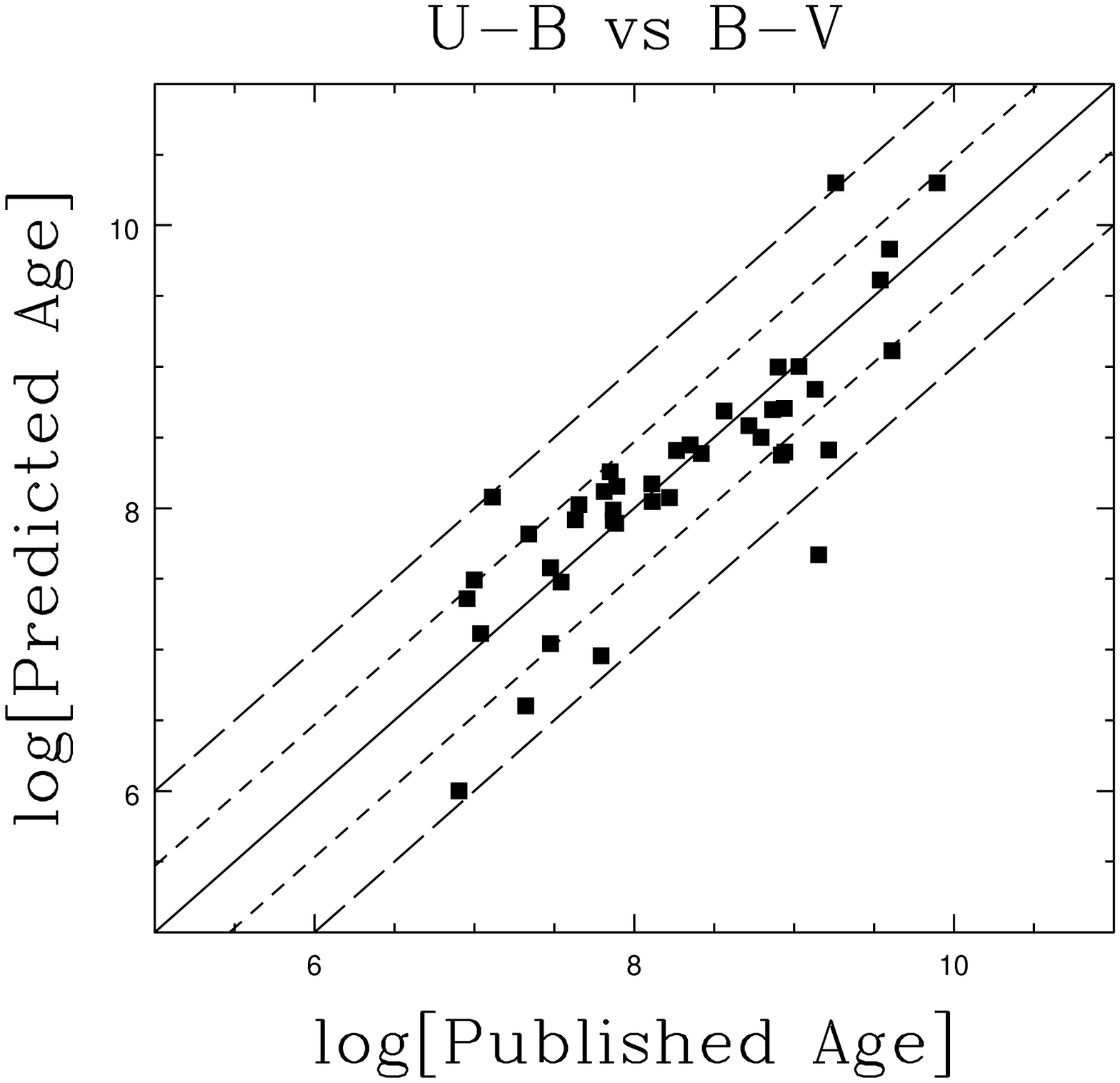, height=7.0cm}}
\centering{\epsfig{figure=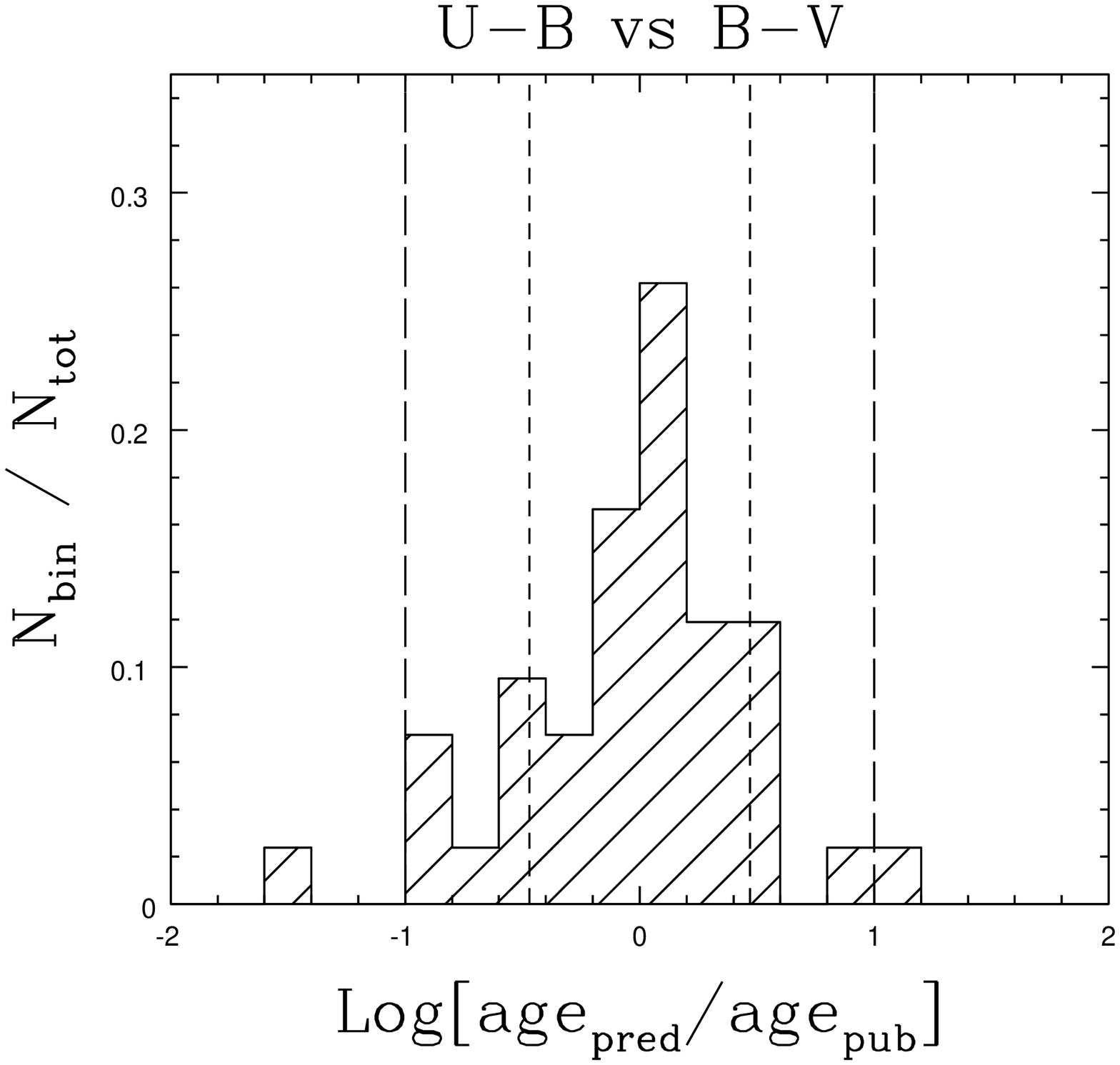, height=7.0cm}}
\centering{\epsfig{figure=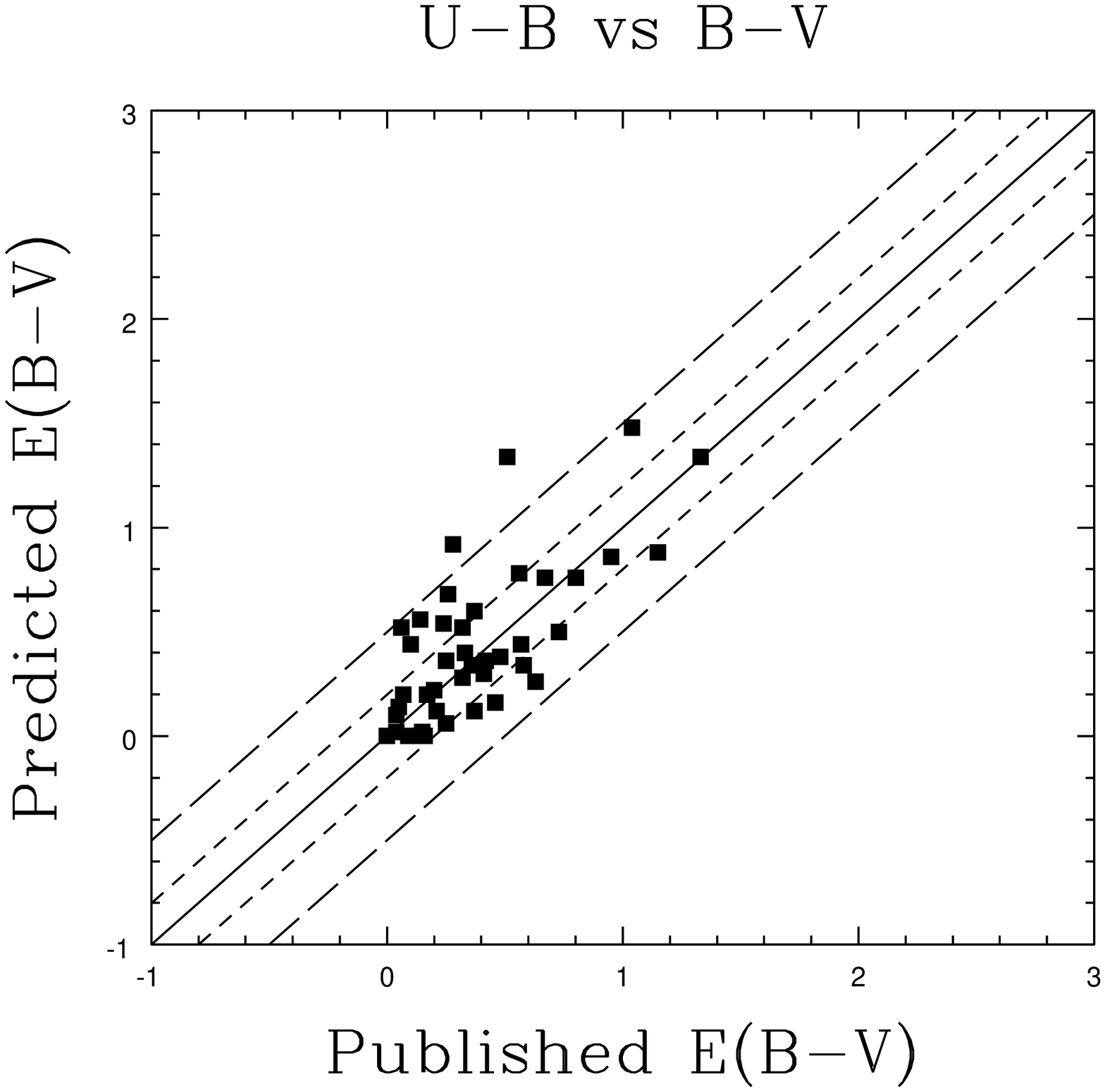, height=7.0cm}}
\centering{\epsfig{figure=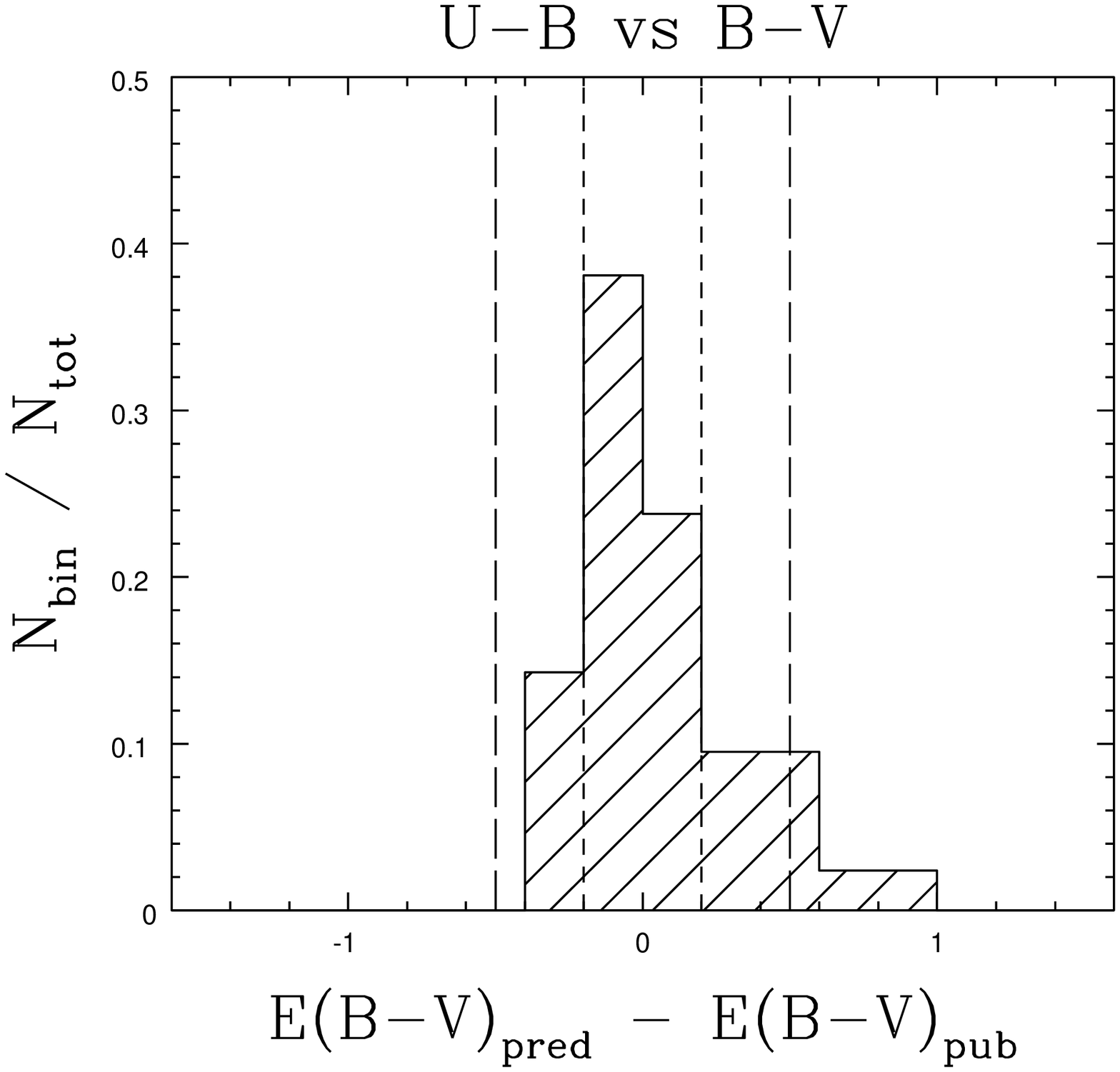, height=7.0cm}}
\caption[f2a,b,c,d]{ Top Left:  Model predicted age versus the published age of the OCs in the WEBDA sample, as determined by the (U$-$B) vs.\ (B$-$V) colours.  The solid black line is y=x, the inner dashed lines represent a factor of 3 difference in age, and the outer long-dashed lines represent a factor of 10 difference in age.  Top Right:  Histogram showing the distribution of the difference between the predicted and published ages.  The inner and outer dashed lines represent a factor of 3 and 10 change in age respectively.  Bottom Left:  Model predicted extinction vs.\ published extinction.  The solid black line is y=x, the inner dashed lines represent a change of 0.2 mag, and the outer long-dashed line represents a change of 0.5 mag.  Bottom Right:   Histogram showing the distribution of the difference between the predicted and published E(B$-$V).  The inner and outer dashed lines represent a change of 0.2 and 0.5 mag respectively. }
\label{f2}
\end{figure*}

\begin{table*}
\centering
\begin{minipage}{140mm}
\caption[fit-quality]{Prediction Quality} 
%\vspace{.2in} 
\begin{tabular}{lrcrlllrrr} 
\hline \hline 
Colours & OCs  & Good     & Rec. & Unc  & Unc & Unc  & w/in 3x  & w/in 3x      & w/in 3x \\ 
Used    &      & Fits$^1$ & Ages$^2$ & down$^2$ & up$^2$  & mean$^2$ & $\odot $ & 0.2 $\odot $ & 2.5 $\odot $ \\ 
\hline 
U$-$B & 44 & 44/44/44  & 98\% & 1.03 & 2.08 & 1.56 & 	45\% & 57\% & 59\% \\ 
B$-$V & 62 & 62/62/62  & 100\% & 1.18 & 2.42 & 1.80 & 	45\% & 45\% & 42\% \\ 
V$-$R & 6 & 6/6/6  & 100\% & 1.37 & 2.87 & 2.12 & 	33\% & 17\% & 50\% \\ 
V$-$I & 15 & 15/15/15  & 100\% & 1.40 & 2.61 & 2.01 & 	47\% & 27\% & 40\% \\ 
U$-$B B$-$V & 42 & 42/42/42  & 98\% & 0.84 & 1.14 & 0.99 & 	83\% & 69\% & 74\% \\ 
U$-$B V$-$R & 5 & 5/4/5  & 100\% & 1.33 & 1.48 & 1.40 & 	20\% & 25\% & 20\% \\ 
U$-$B V$-$I & 11 & 11/8/11  & 100\% & 0.75 & 1.07 & 0.91 & 	45\% & 62\% & 45\% \\ 
B$-$V V$-$R & 6 & 6/6/6  & 100\% & 1.26 & 2.89 & 2.07 & 	17\% & 17\% & 17\% \\ 
B$-$V V$-$I & 15 & 15/15/15  & 87\% & 0.69 & 2.89 & 1.79 & 	67\% & 13\% & 60\% \\ 
V$-$R V$-$I & 5 & 4/4/5  & 100\% & 0.59 & 3.50 & 2.05 & 	75\% & 25\% & 60\% \\ 
U$-$B B$-$V V$-$R & 5 & 5/5/5  & 100\% & 1.16 & 1.65 & 1.40 & 	0\% & 80\% & 20\% \\ 
U$-$B B$-$V V$-$I & 11 & 10/7/11  & 100\% & 0.63 & 1.22 & 0.93 & 	60\% & 57\% & 55\% \\ 
B$-$V V$-$R V$-$I & 5 & 4/4/5  & 100\% & 0.36 & 3.72 & 2.04 & 	50\% & 0\% & 40\% \\ 
U$-$B B$-$V V$-$R V$-$I & 4 & 3/2/3  & 100\% & 0.39 & 1.46 & 0.93 & 	67\% & 50\% & 67\% \\ 
\hline \\ 
%\footnotesize{$^1$ solar/0.2 solar/2.5 solar}\\
\multicolumn{10}{l}{\footnotesize{$^1$ solar/0.2 solar/2.5 solar}}\\
\multicolumn{10}{l}{\footnotesize{$^2$ solar metallicity assumed}}\\
\end{tabular} 
\end{minipage}
\end{table*}

We present Table 2 to discuss both the accuracy and precision of
various colours as age indicators.   Column one is the colour
combination, column two is the number of OCs in the sample with those
colours available, and column three is the number of good fits for the
solar, 0.2 solar and 2.5 solar models, where the number of good fits
is the number of  models in our grid that give $\chi^2\leq$ N.  Column
four is the percentage of recovered ages, which describes how often
the predicted age uncertainties overlapped with the uncertainties in
the measured ages when solar metallicity is assumed.   The results
from comparisons to models of 0.2 solar and 2.5 solar abundances are
similar so are not shown.  Column five lists the average uncertainty
in predicted log(t) in the negative direction for all the OCs, column
six lists the average uncertainty in predicted log(t) in the positive
direction for all the OCs, and column seven lists the average of the
positive and negative uncertainties for all the OCs.   It should be
noted  that the model ages range from a log(t) of 6 to 10.3, therefore
2.15 is the maximum mean positive and negative uncertainty in log(t).
Columns eight, nine, and ten are the percentages of  predicted ages
that are within a factor of 3 of the published ages for the solar
abundance model, the 0.2 solar abundance model, and the 2.5 solar
abundance model, respectively.  This percentage is determined for the
best-fit ages without consideration of the predicted or measured age
uncertainties.  Unfortunately, the only samples of OCs large enough to
draw any significant interpretations are the (U$-$B), (B$-$V) and
(U$-$B) and (B$-$V).  Results from the other  combinations of colours
are merely suggestive.

We see from Table 2 that for several colour combinations the
uncertainties in predicted age are relatively large.  This is because
we have assumed the maximum uncertainty in measured colours for the
OCs.  Using the actual uncertainties in integrated colours for this
sample would not change the predicted ages but would result in better
age constraints.  Unfortunately the uncertainties in integrated
colours of each of these OCs are not available.  However, the relative
predicted age uncertainties provide a good measure of the precision
afforded by each colour and colour combination for age-dating.
Looking at the mean uncertainties (column 7 in Table 2) we see  that
the precision is improved when the (U$-$B) colour is included.

We see from this table that by using the (U$-$B) and (B$-$V) colours
up to 83\% of the predictions are within a factor of 3 of the
published values.   For the best-fit (U$-$B), (B$-$V), and (U$-$B) and
(B$-$V) samples, we have  determined the differences between the
best-fit predicted and measured  log ages.  We find that 68\% of the
differences are less than 0.77 dex, 0.86 dex, and 0.44 dex for the
(U$-$B), (B$-$V), and (U$-$B) and (B$-$V)  samples respectively.  This
indicates that (U$-$B) alone gives more accurate ages than (B$-$V)
alone.  We also see that combining (U$-$B) with (B$-$V) further
improves the accuracy of the predicted ages.

Table 2 shows the importance of the (U$-$B) colour in measuring the
age of a cluster.  To the extent that our sample size allows such
comparisons,  combinations of colours that include (U$-$B) tend to
have more accurate and more precise age predictions.   The fact that
single colours do not do a good job of constraining the ages is likely
because there are three degenerate unknowns, age, reddening, and
metallicity, and two filters are not enough to  break these
degeneracies.  This is consistent with results from \citet{and04}.
However, the (U$-$B) colour alone is a better predictor of age than
any combination of colours that do not include (U$-$B), regardless of
the assumptions about metallicity, consistent with results from
\citet{gil02} and \citet{and04}.  The (U$-$B) colour reflects the
contributions from the young and intermediate age stars.  Tracing the
younger stars in a population is critical to constrain the age.
Colours that do not include {\it U} only follow the slower-evolving
stars and are not as sensitive to changes in age.

\subsection{Extinction and Metallicity}

The extinction is predicted well by the same colours that predict the
age well.  For example, using the (U$-$B) and (B$-$V) colours
we find that 62\% of the \ebv\ estimates are within 0.2 mag of the
published values.  The colour that is the most sensitive to age,
(U$-$B), is once again critical.  

We get similar numbers of good fits to the cluster ages whether we
assume 0.2 solar, solar, or 2.5 solar abundances (see Table 2).  This
implies that, at least within the metallicity range of our sample OCs,
knowing the metallicity is not critical for getting a reasonable fit
to the cluster age.  Note that we are not fitting for the
metallicities of these clusters, but instead are testing to see how
well we can recover the correct ages and extinctions with standard
solar and near solar metallicity models.

\begin{figure}
\centering{\epsfig{figure=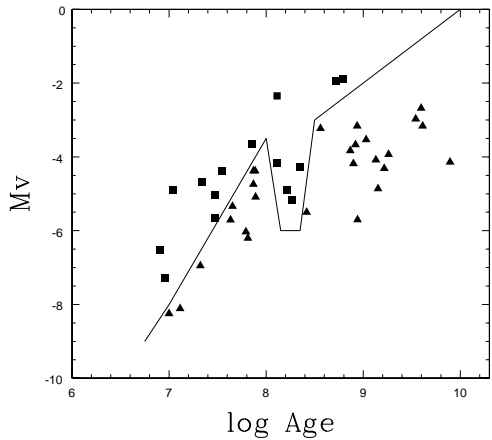, height=7.0cm}}
\caption[f3.eps]{Published M$_{V}$ plotted against the published ages of the open clusters in our sample.  The curve approximates the lowest luminosity limit curve in \citet{cer04}.  The filled black squares are the OCs fainter than the curve while the filled black triangles are brighter. \label{f3}}
\end{figure}

\subsection{The Effect of H$\alpha$}

The SB99 model colours do not include nebular H$\alpha$ emission.
This does not impact our analysis of the integrated colours of the OCs
because the stellar components are resolved and no H$\alpha$ is 
included in either the models or the data.  However, this should be
considered when models are compared to young unresolved clusters
in other galaxies.

For clusters younger than $\sim$10 Myr, the H$\alpha$ emission
contributes significantly to the Johnson {\it R} band and minimally to the
Johnson {\it V} band.   From SB99, it is seen that a 1 Myr model has an
EW(H$\alpha$) of $\sim$2500 \AA, a 5 Myr model has an EW(H$\alpha$)
of $\sim$400 \AA, and by 10 Myr the  EW(H$\alpha$) is $\sim$10 \AA.
For comparisons to young unresolved clusters, H$\alpha$ emission
should be included in the model SEDs.  For a discussion about adding
H$\alpha$ emission to models to compare to the clusters in Arp 285, 
see \citet{smi08}.

\subsection{The Effect of Stochastic Sampling of the IMF}

It has been shown that traditional synthesis models usually return the
mean value of the integrated stellar population distribution
(e.g. \citealp{cer06,lur06}).   While this is correct on average, it
is not necessarily correct in individual cases because the mean may not be
representative of the actual values.  Galactic open clusters are
typically low mass so will be subject to the effects of stochastic
sampling of the IMF (e.g. \citealp{cer03}).

\begin{figure}
\centering{\epsfig{figure=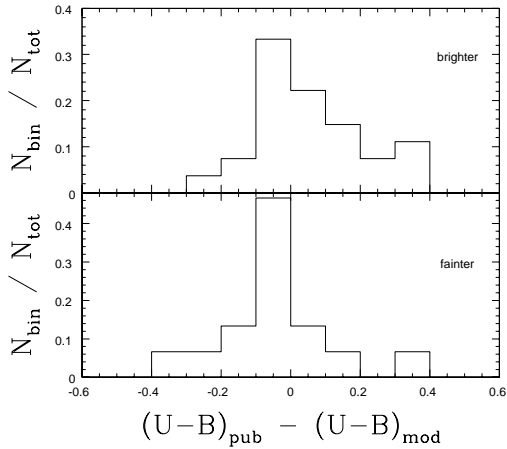, height=7.0cm}}
\centering{\epsfig{figure=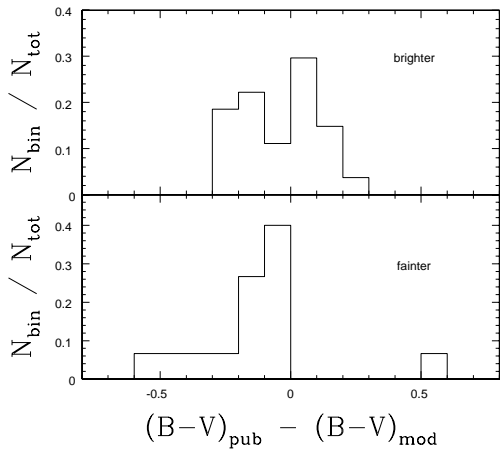, height=7.0cm}}
\caption[f4a,b]{Top:  Distribution of the difference between the published (U$-$B)$_{pub}$ and the model (U$-$B)$_{mod}$ for the open clusters in our sample.  The model colours assume the published ages.  The colours are reddening-corrected.  The histograms are divided into two groups;  fainter than and brighter than the lowest luminosity limit curve, as shown in Figure 3.  Bottom:  Same as above but for the (B$-$V) colours. \label{f4}}
\end{figure}

\begin{figure}
\centering{\epsfig{figure=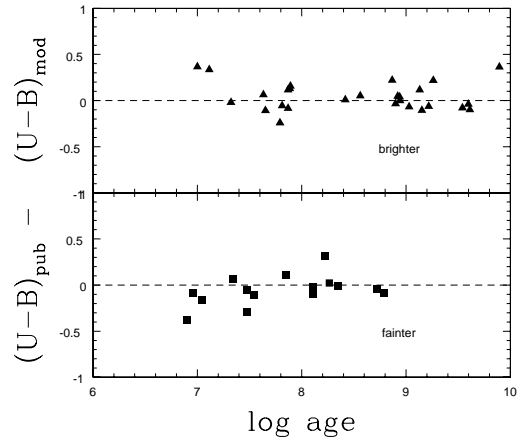, height=7.0cm}}
\centering{\epsfig{figure=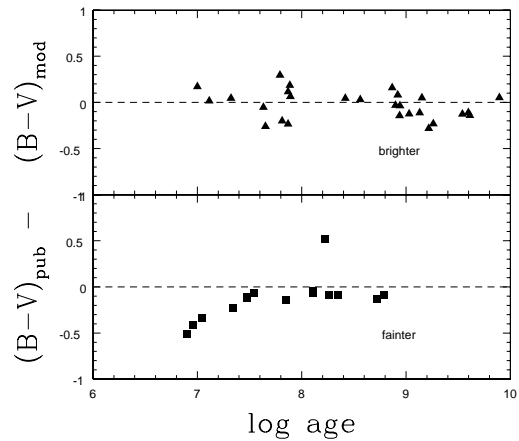, height=7.0cm}}
\caption[f5a,b]{Top: Difference between the published and model (U$-$B) colours against the published ages of the clusters.  Bottom:   Difference between the published and model (B$-$V) colours against the published ages of the clusters.  The two panels describe the same two groups as above.  The dashed lines show zero colour difference. \label{f5}}
\end{figure}

\citet{cer04} define the lowest luminosity limit as a criterion that
must be met in order to compare an observed cluster to a model
cluster.  The total luminosity of the observed cluster must be larger
than the  individual contribution of any of the stars in the model.
According to \citet{cer04}, clusters brighter than this limit may
or may not have a well-sampled IMF, but clusters fainter than this
limit will be misrepresented by synthesis models.

Figure 3 is a plot of the published M$_{V}$ plotted against the
published ages of the OCs in our sample.   The curve approximates the
lowest luminosity limit curve in \citet{cer04}.   The filled squares
are the OCs fainter than the curve, while the filled triangles are OCs
that are brighter.  There are 15 OCs fainter and 27 OCs brighter
than the curve.  From this figure it can be seen that several of
the OCs in our sample  are brighter than the lowest luminosity limit,
but the majority are  fainter.  This suggests that many clusters in
our sample (those fainter) should show these stochastic effects, while
several might not (those brighter).

Figure 4 shows the distributions of the differences between the
published and the model (U$-$B) colours and the published and model
(B$-$V) colours for the OCs in our sample.  The measured colours are
reddening-corrected.  The model colours assume the published ages.  To
make comparisons, only OCs with both integrated  (U$-$B) and (B$-$V)
colours are included in these distributions.  The histograms are
divided into 2 groups, those brighter and those fainter than the
lowest luminosity curve, as in Figure 3.

For the (U$-$B) colour, a Kolmogorov-Smirnov (K-S) test cannot rule out
that the brighter and fainter samples originated from the same parent
distributions.  The rms for the differences in (U$-$B) are 0.16 and
0.17 for the brighter and fainter samples respectively.  However, for
(B$-$V), a K-S test suggests a probability of only 0.012 that the two
samples (brighter and fainter) originate from the same  parent
distribution.  The rms values are 0.15 and 0.25 for  the (B$-$V)
distributions respectively.
The scatter appears slightly larger for the fainter samples,
with the (B$-$V) difference being slightly skewed to negative
numbers (see Figure 4).

Figure 5 shows the differences between the published and model (U$-$B)
and (B$-$V) colours plotted against the published ages using the same
sample as in Figure 4.  The measured colours are reddening-corrected.
For each colour difference the OCs are divided into the same bins as
described above.  From Figure 5 it can be seen that the scatter in the
(U$-$B) colour differences do not appear to be correlated with age.
The fainter (B$-$V) difference sample appears to have a slight
dependence on age in that the published (B$-$V) colours are bluer than
the model colours for the younger OCs,  consistent with results in
\citet{bru02} (see below).   Given this, the K-S results, and the
larger rms, we might be seeing stochastic sampling effects in the
differences between the measured and model (B$-$V) colours.

In Figure 6, we plot the published reddening-corrected (U$-$B) colour
against the published reddening-corrected (B$-$V) colour for our OCs.
This figure is divided into the same two groups described above.  Also
on this figure are our SB99 models with E(B$-$V) of 0.0 mag (solid
curve) and 0.5 mag (dashed curve).  This figure shows that the scatter
of these colours around the models is not a function of M$_{V}$.  
Note that relatively young clusters (that is, blue clusters)
tend to lie above and to the left of the zero extinction model curve.
Figures 4 and 5 show that this is predominately due to the fainter clusters
being bluer in (B$-$V) than the models.

\begin{figure}
\centering{\epsfig{figure=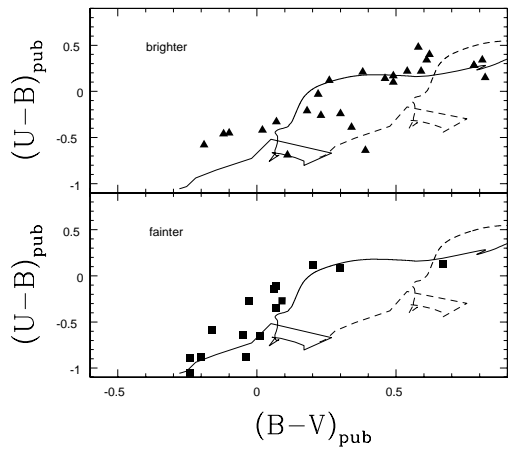, height=7.0cm}}
\caption[f6]{Published (U$-$B) against published (B$-$V) for the OCs.  The figures are divided into 2 groups as described above.  The curves are our SB99 models with E(B$-$V) of 0.0 mag (solid curve) and 0.5 mag (dashed curve). \label{f6}}
\end{figure}

Our results can be compared to population synthesis studies that
explicitly investigate stochastic effects.  For example, \citet{bru02}
and \citet{bru03} have run Monte Carlo simulations of low mass star
clusters, to investigate how stochastic effects can cause observed
colours to differ from models with full-sampled IMFs.  In their
simulations, young model clusters with masses of 10$^3$ M$_{\sun}$
also lie above models with fully-sampled IMFs on a (U$-$B) vs. (B$-$V)
plot \citep{bru02}, consistent with what we see in Figure 6.   Thus at
least some of the observed scatter may be due to stochastic effects.

From these figures and tables we see suggestive evidence of stochastic
sampling of the cluster IMF in this sample of OCs.   It is possible
that the uncertainties in the measured colours and/or ages are large
enough (0.14 mag and 19\% respectively) that the effects of stochastic
sampling are washed out.   However, figure 1 in \citet{cer03} shows
that the expected scatter in the integrated (B$-$V) of a 10 Myr
cluster with $10^{3}$ or less stars is much larger than our assumed
colour uncertainties.  Most of the OCs in our sample are much  older
than 10 Myr.  It is possible that stochastic sampling effects are
smaller in (B$-$V) for older clusters.  After $\sim10$ Myr the most
luminous stars will have evolved off of the main sequence and will
contribute more strongly to the red end of the integrated cluster SED.
\citet{deg05} remark that stochastic sampling effects affect broadband
photometry to a smaller extent than spectroscopy.  In a follow up
paper (Hancock et al., in prep.)  we will explore a more detailed
analysis of the stellar content of these resolved OCs to further
investigate the extent to which stochastic sampling of the IMF affects
the integrated broadband colours.  If stochastic sampling effects are
observationally characterized in  this sample of OCs, they will
provide a good bench mark for testing  future population synthesis
models.

\subsection{The Effect of Cluster Dissolution}

Current stellar population synthesis models do not take into account
the effects of cluster dissolution.  Mass segregation causes the
massive stars to concentrate towards the centre of a cluster while the
low-mass stars tend to populate the outer regions.   A consequence of
mass segregation is that low-mass stars tend to be preferentially
ejected from star clusters.  The mass function (MF) therefore changes
drastically during  the evolution of a cluster, as a consequence,
becoming less steep  than the IMF (e.g. \citealp{bau03}).  The
preferential loss of low-mass stars will change the integrated colours
of a cluster (e.g. \citealp{lam06a}).  Such a time evolution of the
integrated colours resulting from cluster dissolution can be difficult
to disentangle from stellar evolutionary effects.

The colour change of a star cluster due to dissolution depends
upon its age, its mass, and its dissolution time scale (t$_{dis}$), 
with the effect being larger for later ages and longer t$_{dis}$.
\citet{lam06a} show that, for a $10^5$ M$_{\odot}$ cluster, the 
(B$-$V), (V$-$I) and (V$-$K) colours are not significantly affected by
the  preferential loss of low-mass stars during the first 80\% of
their $t_{dis}$, with $\Delta$(V$-$I)$=0.03$ and 0.1 mag for a
t$_{dis}$ of 1 Gyr and 10 Gyr respectively.  However, in the
last 20\% of a cluster's life, the colour effect can be quite
large, up to $\Delta$(V$-$I)$=0.3$ mag.

Our sample clusters likely have masses between $10^3$ and $10^4$
M$_{\odot}$.  This implies t$_{dis}$ between $\sim$100 Myr and 2 Gyr
\citep{lam06b}.   Given that 45\% of our sample have ages less than 80
Myr and  86\% have ages less than 1.6 Gyr (see Table 1), it is likely
that most of our clusters are not in the last 20\% of their lifetimes,
but we can not rule out the possibility that some might be near the end
of their lifetimes.  Thus the effect of dissolution on their colours
is likely small, typically smaller than our assumed uncertainties of
0.14 magnitudes.

Using the stellar data available on WEBDA for a  sample of our OCs
covering a representative range of distances, we found that $\sim85\%$
contained stars $\la$0.5 M$_{\odot}$.   To determine the effects that
low mass stars have on the model integrated colours we compared 
Kroupa models with lower mass cutoff of 0.1 M$_{\odot}$ with those
with 0.5 M$_{\odot}$.  We found no significant differences in
the (U$-$B), (B$-$V), (V$-$R), or (V$-$I) colours.

\section{SUMMARY}

We present an empirical assessment of the use of broadband colours as
age indicators for unresolved extragalactic clusters and investigate
stochastic sampling effects on integrated colours.  The population
synthesis code {\it Starburst99} \citep{lei99} and four optical
colours were used to estimate how well we can recover the ages of 62
well-studied Galactic open clusters with published ages.  We conclude
the following:

1)  Galactic open clusters can be used for testing the integrated
properties from population synthesis and serve as reasonable
benchmarks for future assessments of age-dating methods.

2)  The (U$-$B) colour is critical.  Only colour combinations that
included (U$-$B) resulted in good age and extinction predictions,
consistent with previous results.

3)  Only with (U$-$B) included were the predicted age uncertainties
reasonably constrained.

4)  Changes of a factor of $\sim$2 in assumed metal abundance do not
result in significantly different predictions of cluster age.  This
indicates that the uncertainties in predicting the age of a cluster
resulting from the age-reddening degeneracy dominate over the other
sources of degeneracy in these optical bands, over our age range
and metallicity range.  Another possibility is that uncertainties in 
the measured colours (and hence the predicted ages) dwarf the 
metallicity-age degeneracy. 

5)  A $\chi^2$ minimization and a $\Delta\chi^2$ defined to give 68\%
confidence levels provide reliable age estimates, and more
importantly, reliable age uncertainties.  The difference in the
photometric ages and the ages derived from the HR-diagrams of our
selected cluster sample with both the (U$-$B) and (B$-$V) colours are
smaller than 0.5 dex for 79\% of the clusters, with a maximum
difference of 1.5 dex.

6)  It is likely that the uncertainties in the measured colours and/or
ages are large enough (0.14 mag and $\sim20\%$ respectively) that the
effects of stochastic sampling are washed out.  A more detailed
analysis of the stellar content of these resolved OCs will be required
to further investigate the extent to which stochastic sampling of the
IMF affects the integrated broadband colours.

7)  If stochastic sampling effects are observationally characterized
in  this sample of OCs, they will provide a good bench mark for testing
future population synthesis models.

\section*{Acknowledgments}

We thank the anonymous referee for his/her helpful comments and
suggestions.  This research has made use of the WEBDA database,
operated at the Institute for Astronomy of the University of Vienna.
The authors would like to thank Ernst Paunzen for providing ASCII
tables of the WEBDA data.  The authors also thank Alessandra Stone for
help with the data acquisition.  This work has been supported by the
NASA LTSA  grant NAG5-13079.

\bsp

\label{lastpage}

\end{document}